\documentstyle[epsf,epsfig,rotate]{mn}

\newcommand{\au}     {\hbox{$\,\mathrm{au}$}}

\newcommand{\kmps}   {\hbox{$\,\mathrm{km}\,\mathrm{s^{-1}}$}}
\newcommand{\myr}    {\hbox{$\,\mathrm{Myr}$}}

\newcommand{\yr}     {\hbox{$\,\mathrm{yr}$}}

\begin{document}

\title[Close stellar encounters with planetesimal discs]
{Close stellar encounters with planetesimal discs:\\
The dynamics of asymmetry in the $\beta$ Pictoris system}

\author[J.D. Larwood and P.G. Kalas]
{J.D. Larwood$^{1}$\thanks{E-mail: j.d.larwood@qmw.ac.uk} and
P.G. Kalas$^{2}$\thanks{E-mail: kalas@astron.berkeley.edu}\\
$^{1}$
Astronomy Unit, Queen Mary \& Westfield College, Mile End Road,
London E1 4NS\\
$^{2}$
Astronomy Department, University of California, 601 Campbell Hall,
Berkeley, CA 94709, USA}

\date{Accepted 2000 November 14}
\volume{000}
\pagerange{\pageref{firstpage}--\pageref{lastpage}}
\pubyear{0000}

\maketitle
\label{firstpage}

\begin{abstract}
We numerically investigate the dynamics of how a close stellar
fly-by encounter of a symmetrical circumstellar planetesimal disc
can give rise to the many kinds of asymmetries and substructures
attributed to the edge-on dusty disc of $\beta$ Pic.
In addition we present new optical coronagraphic observations of
the outer parts of $\beta$ Pic's disc, and report that
the radial extent is significantly greater
than was found in previous measurements.
The northeasterly extension of the disc's midplane is now
measured out to $1835$\au~from the star; the southwesterly
component is measured out to $1450$\au. Thus the asymmetry
in the length of the former with respect to the latter is
approximately $25$ per cent.
We proceed to use the length asymmetry induced in the distribution of
simulation test particles as the principal diagnostic feature when
modelling the disc response in order to constrain fly-by parameters.
In particular we favour a low inclination prograde and near-parabolic
orbit perturber of mass approximately $0.5$\,M$_\odot$.
These initial conditions suggest that the perturber could have been
physically associated with $\beta$ Pic prior to the encounter. Thus
we also consider the possibility that the perturber could be bound
to $\beta$ Pic: a consideration also of general interest where dust discs
are known to exist in binary star systems.
We show that a further consequence of a low velocity encounter is that
the perturber could have captured planetesimals from the
$\beta$ Pic disc, and we deduce that as a result of this the
perturber could display a dust disc that is presently amenable to observation.
In some of our models,
we can relate groupings of perturbed particles to the
large-scale structure of the $\beta$ Pic disc.
The groupings correspond to:
high eccentricity and inclination particles that
reach apocentre and maximum height in the southwest,
moderately eccentric and low inclination particles that reach
apocentre in the northeast,
and relatively unperturbed particles inside $\sim$200\au~radius.
\end{abstract}


\section{Introduction}

The starlight of $\beta$ Pic scatters off dust grains in its
edge-on circumstellar disc (Smith \& Terrile 1984). Optical
imaging reveals that the disc exhibits asymmetrical structure
with respect an ideal disc on projected radial scales from
$\sim$50\au~(Burrows et al. 1995, Mouillet et al. 1997) to
$\sim$1000\au~(Smith \& Terrile 1987, Kalas \& Jewitt 1995). 
Kalas \& Jewitt (1995) presented sensitive $R$-band images
that showed the northeast (NE) extension could be detected
out to $790$\au, whereas the southwest (SW) extension was
detectable out to just $650$\au, corresponding to a
$\sim$20 per cent length asymmetry.
More recently, the longer NE midplane extension has been discovered
to carry several brightness enhancements from
$\sim$500 to $800$\au~(Kalas et al. 2000; hereinafter KLSS),
that have been attributed to a series of rings within the disc's
midplane. Similar features in the SW extension were not detected
and so eccentric rings were hypothesised.

KLSS presented a dynamical model in which the stellar fly-by
scenario (Kalas \& Jewitt 1995) for
generating the disc asymmetries was applied to the $\beta$ Pic system.
Their finding was that the asymmetry types present in the $\beta$ Pic
disc have analogues in the dynamical response of a particle disc
to a perturbing stellar fly-by encounter. Furthermore, they
demonstrated that these features could occur simultaneously with
the formation of eccentric circumstellar rings. A simple analysis of the
perturbed disc model was found to be in good quantitative agreement
with the observations of both $\beta$ Pic's disc asymmetries and the midplane
features; for particular values of the viewing angle and epoch, and
fly-by parameters. In this paper we present new observations and measurements
of the length asymmetry of $\beta$ Pic's scattered-light disc,
and then go on to simulate the dynamics of a stellar flyby
encounter with an initially symmetrical disc of particles.

Similar numerical modelling to that presented here and in KLSS
has been considered previously in the context of
angular momentum transfer between the perturber's orbit and a
protostellar disc (Clarke \& Pringle 1993, Hall, Clarke \& Pringle 1996),
the dynamics of the Edgeworth-Kuiper belt (Ida, Larwood \& Burkert 2000),
and the tidal interaction of spiral galaxies (Toomre \& Toomre 1972).
Here we shall consider a wider range of encounter velocities
(from bound orbits through to hyperbolic trajectories)
and stellar masses (from M dwarfs to A stars, when scaled to the
mass of $\beta$ Pic) than has been studied before in any context.
We also include a discussion of the orbital dynamics of
captured particles with respect to the perturber. This paper
can be considered as a companion to KLSS since here we also supplement
their finding of circumstellar ring formation in similar simulations
with an explanation of the dynamical origin of the features.

As in other works (Whitmire, Matese \& Tomley 1988,
Mouillet et al. 1997, KLSS)
we assume that the simulation particles' distribution represents
the distribution
of an underlying disc of planetesimals, which are parent bodies
whose infrequent and therefore dynamically inconsequential collisions
supply the dust scattering surface that is observed in the
real system. A more complete model should include consideration
of grain removal and generation processes in determining the
appearence of the perturbed disc (reviewed by Backman \& Paresce 1993).
However, we defer treatment of those issues to future work,
and focus here on the first-stage problem of the dynamics.
This stage is important in deducing the mass and orbital parameters
of the postulated stellar perturber, for which it is possible to perform
star catalogue searches
(e.g. Kalas, Deltorn \& Larwood 2000), and we proceed to do this by
examining the length asymmetry, and other measures of the disc
response, as a function of stellar encounter parameters. However, we choose
to examine the length asymmetry as the principal diagnostic measure
of the disc response since it is the most distinct such feature
both in the observations and the dynamical models. In our future work
we shall also consider the subtler asymmetries in greater detail.

In Section $2$ we present deep optical coronagraphic images
of the disc. In Section $3$ we describe the results of our
numerical investigations of the dynamics of tidally disrupted
particle discs, and provide general discussion of the induced
asymmetries and particle stripping. In Section $4$ we discuss
our findings and summarise the conclusions.

\section{New observational results}

New optical coronagraphic observations of the environs of $\beta$ Pic were
obtained with the University of Hawaii $2.2$\,m telescope on
2000 January 31. A key goal was to maximize sensitivity to the
outer regions of the disc. The coronagraph used has 12 interchangeable,
circular, focal-plane occulting masks,
from which we selected a mask $12.5$\,arcsec
in diameter. This particular mask size is approximately
two times larger than the minimum size allowable by the 
seeing conditions, and it partially obscures the image out to
$120$\au~projected radius about $\beta$ Pic. However, it also minimizes the 
scattered light in the field that dominates the background noise.
Additionally, the occulting mask increases the efficiency of the
observations by maximizing the integration time permitted before
CCD saturation occurs just beyond its edge.

\begin{figure}
\centerline{\epsfig{file=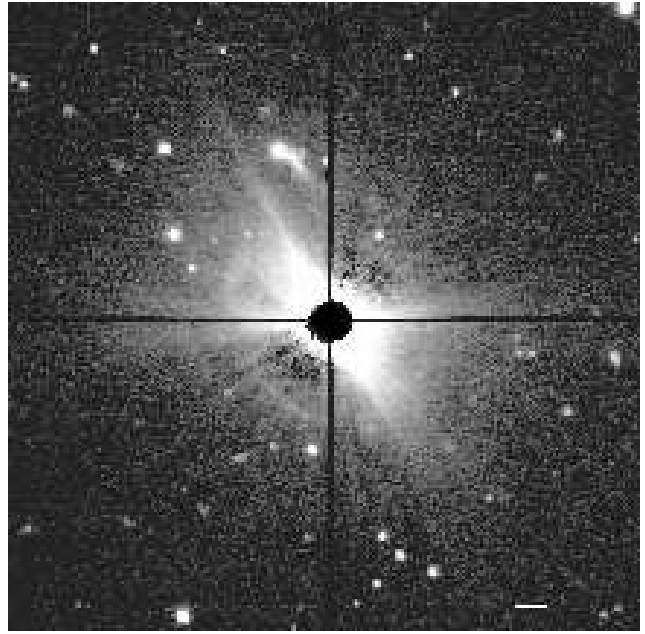,width=84mm,angle=0}}
\caption{PSF-subtracted, $R$-band coronagraphic image
centred on $\beta$ Pic. North is up and east is left.
Bars in the northeast and southwest 
quadrants represent 10\,arcsec and are placed
103.6\,arcsec (2000\au) radius from $\beta$
Pic. The NE disc midplane is detected
near $2000$\au~radius, whereas the SW midplane
is traced to approximately $1000$\au~radius.
Other large scale extended features in
this image are due to instrumental scattered light 
(e.g. radial linear features, a narrow ring segment
in the southeast quadrant, and amorphous 
structure near a bright star north of $\beta$ Pic).}
\label{fig1}
\end{figure}
\begin{figure}
\centerline{\epsfig{file=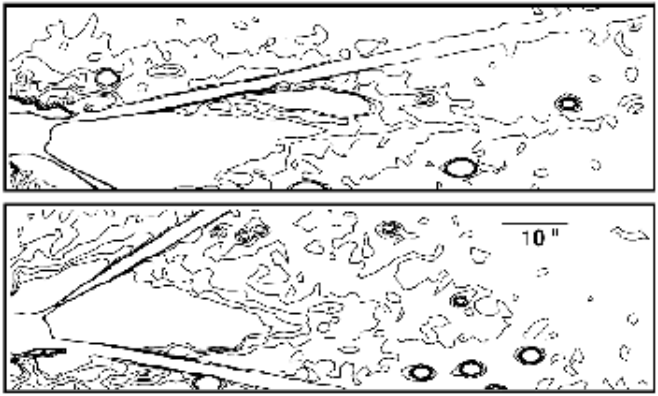,width=87mm,angle=0}}
\caption{Surface brightness contours for the 
NE midplane (top) and SW midplane (bottom).
The original image (Fig. 1) has been rotated
60\degr~anticlockwise such that the
midplane lies along pixel rows. This
image is separated along the vertical axis
and the NE midplane then reflected
such that the two sections of midplane point in the
same direction. The outer contour
represents 25.0\,mag\,arcsec$^{-2}$ and
has radial extent approximately 25 per cent greater for
the NE extension than for the SW extension. The binning
procedure for the data is described in the text.}
\label{fig2}
\end{figure}
\begin{figure}
\centerline{\epsfig{file=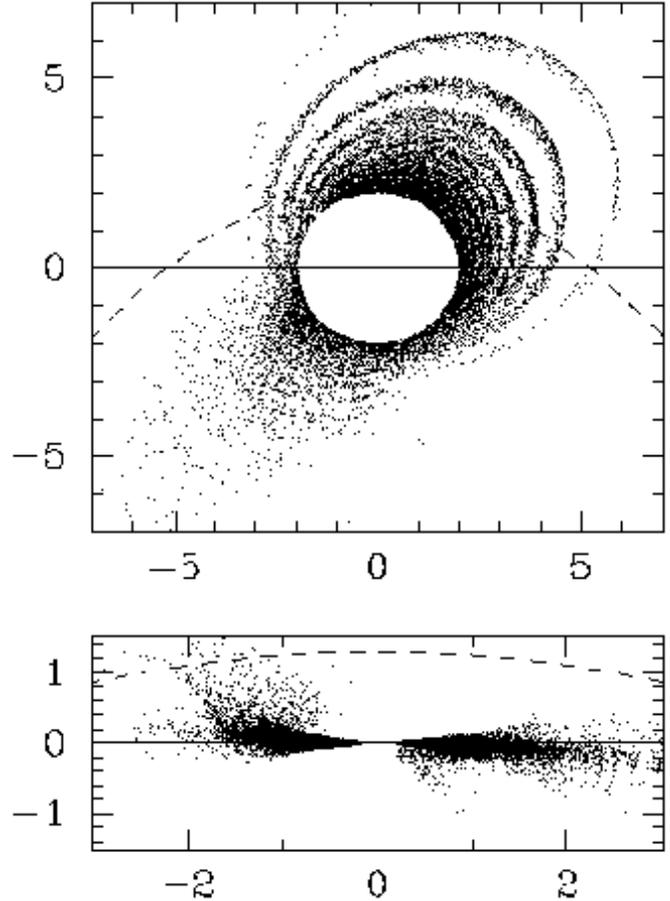,width=90mm}}
\caption{The face-on and edge-on views of a sample of simulation particles
in the computational frame centered on the primary mass (labelled
Fig.$2$ in KLSS).
The primary is of unit mass and $G=1$, giving a time unit such that
the circular orbit period at unit radius is $2\pi$.
The horizontal axes correspond to the line of nodes of the perturber's
orbit. Parameters are: perturber mass of $0.3$,
parabolic pericentre distance of $2.6$,
pericentre inclination of $30\degr$. The disc of $\sim$10$^6$ particles
is initially distributed between $0.2$ and $2$ radius units.
The initial vertical density
is exponentially decreasing, with an outer height of $0.6$,
and the aspect ratio and the radial surface density follow
radial power laws of index $-1.5$. A random sample of $\sim$25 per cent
of the particles outside
the initial radius are shown in the top frame, the bottom frame shows
a thin slice of particles about the horizontal axis.
The path of the perturber is indicated with a
dashed line and it enters the frame from the right-hand side.
The particles' positions and velocities are integrated with a second-order
leapfrog scheme; the perturber orbit is prescribed as a function of
time. The configuration shown is for a time of $160$ units after
pericentre passage. The perturber was introduced at a separation from
the primary of $20$ units.}
\label{fig3}
\end{figure}

We obtained 122 integrations of 30 seconds each through
an $R$-band filter ($\lambda\sim$0.67\,$\mu$m). Re-imaging
optics provided a pixel scale of $0.407$\,arcsec\,pixel$^{-1}$
on a $2048\times2048$ CCD.  
The field was masked by the coronagraphic focal plane 
assembly which has a circular clear aperture with diameter
$336$\,arcsec. The airmass
of observation ($>$3.0) combined with high winds, produced
image quality varying from 1.0 to 1.8\,arcsec.
For final data analysis, we selected 77 frames with
the image quality characterized by stellar full-width at
half-maximum (FWHM) $\le1.2$\,arcsec.

Data were bias-subtracted, flat-fielded, and sky-subtracted using
the median sky value in the corners of the 
field for each frame. Comparison stars were also
observed for the purpose of subtracting the 
$\beta$ Pic PSF. However, this subtraction step adds
noise and so we chose to remove the background using
an artificial point-spread function (PSF). The artificial
PSF is a figure of rotation for a seventh-order polynomial
that is a least squares fit to $\beta$ Pic's PSF
sampled in radial directions perpendicular to the disc midplane.

Fig. $1$ shows the final image of $\beta$ Pic with
effective integration time 2310 seconds. The 3$\sigma$
background noise corresponds to a sensitivity of
$R=+25.8$\,mag. The midplane of the SW extension is difficult
to trace visually as a distinct linear feature
beyond $\sim$55\,arcsec ($1062$\au), whereas the NE midplane
is traceable at up to $\sim$95\,arcsec ($1835$\au) radius from $\beta$ Pic.
By these measurements the length asymmetry is approximately $70$ per cent.

A key problem in comparing the lengths of the
two extensions is that the morphology
of the SW extension departs significantly from a linear
feature, becoming vertically distended
and amorphous at large radii. To compensate for this difference in vertical
structure we rotate the image such that the disk midplane lies
along pixel rows and bin the data in the vertical
direction. We do this by summing the light at each pixel location
with the light from its two vertically-adjacent pixels.
This vertical binning procedure smoothes the data and
equalizes the NE and SW FWHM measurements when taking vertical
cuts at given projected radii along the midplane extensions.
Smoothed contours of the resulting image are shown in Fig. $2$.
Here, the 25.0\,mag\,arcec$^{-2}$ contour traces
the NE extension to 95\,arcsec radius, and the SW
disk extension is traced to $\sim$75\,arcsec ($1450$\au).
The length asymmetry by this measurement is therefore
$\sim$25 per cent, which is fractionally larger than the length asymmetry
measured by Kalas \& Jewitt (1995) at about a half of the radial
scale probed here. The asymmetric vertical morphology evident in the
Fig. $2$ contours also confirms the \emph{butterfly asymmetry}
(see Section $3.3$) reported by Kalas \& Jewitt (1995).

\begin{figure*}
\centerline{\epsfig{file=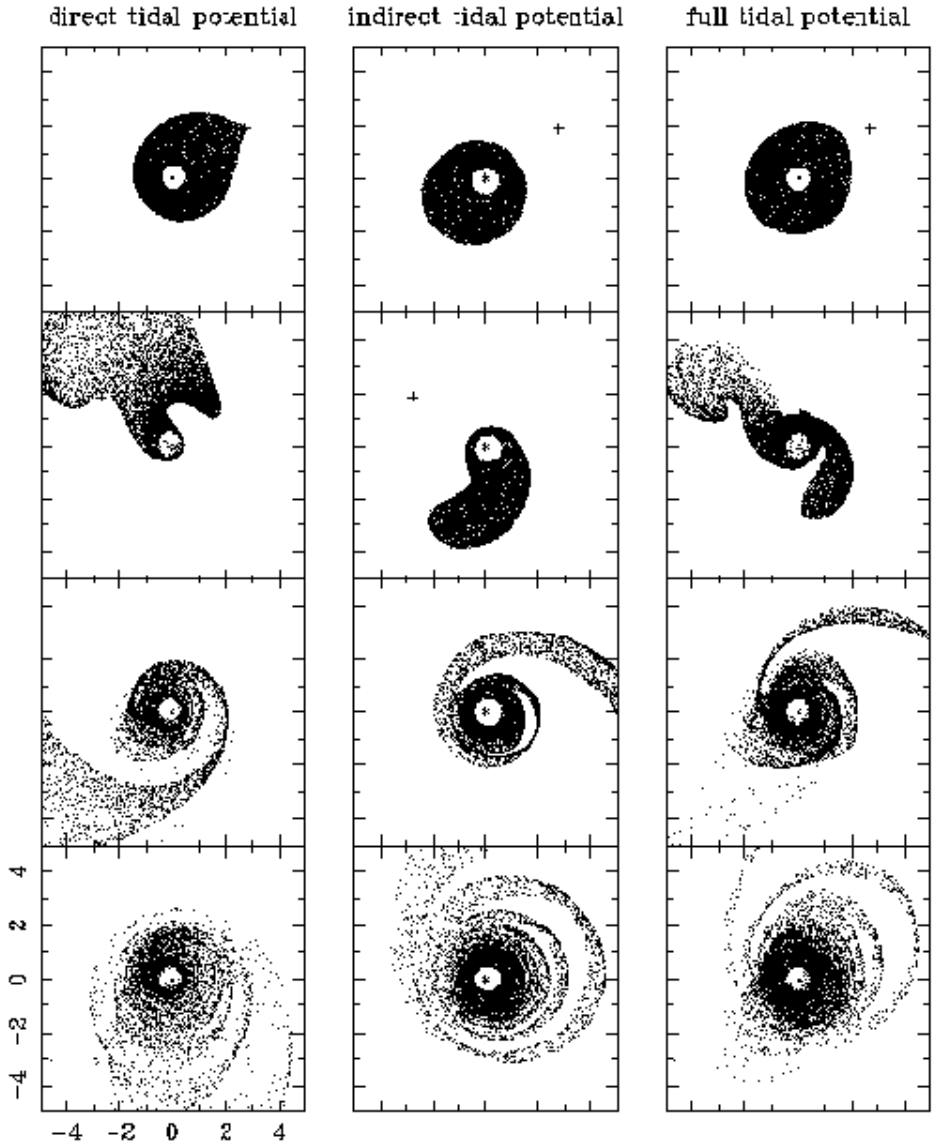,width=130mm}}
\caption{Particle positions at various times for parabolic fly-by encounter
simulations that use different components of the tidal potential.
From the top row of frames to the bottom the times given are
$-3$, $+3$, $+20$, and $+60$. The perturber is plotted with a
cross, the primary star (origin) is plotted with an asterisk.
In these simulations we used $\epsilon=10^{-7}$, $R_{\rm p}=0.1$,
and $R_{\rm s}=0.05$, and removed particles that went beyond radii
greater than 10 from the primary.}
\label{fig4}
\end{figure*}

	\subsection{Summary}

The new data reported above show that the $\beta$ Pic 
disk has a significantly greater radial extent ($1835$\au~for the NE
disk extension; $1450$\au~for the SW) than has been previously measured.
Smith \& Terrile (1987) reported the maximum extent of $\beta$ Pic's disc
at $1300$\au~(i.e 67\,arcsec) for the NE midplane at optical wavelengths.
However, this was known as a sensitivity-limited
value rather than the true physical extent of the disc. Here too
we are unable to define the outer disc extents above the level of the
background noise.

The existence of a disc miplane out to $\sim$10$^3$\au~radius, as well as the
$\sim$10$^2$\au~vertical thickness of the SW extension
(see Fig. $4$ in Kalas \& Jewitt 1995),
is unexpected given our current understanding of planetesimal
formation as applied to the Solar System (reviewed by
Weidenschilling \& Cuzzi 1993). Thus we must conclude that
planetesimals were created at much smaller radii than they are presently
inferred to exist in $\beta$ Pic, and subsequently suffered redistribution.
Moreover this redistribution must allow for the existence of a
lopsided and vertically flared disc.
Global redistribution of this kind is a general outcome
of a close stellar fly-by encounter (KLSS). Below we shall present
numerical investigations of the dynamics involved.

\section{Dynamical models}

In the fly-by model presented in KLSS a transient system
of circumstellar eccentric rings was found to form in the perturbed
disc of particles, as shown in Fig. $3$. Fly-by parameters, and viewing
angle and epoch, were chosen to give the most satisfactory fit
possible to the positions of the midplane brightness peaks found in
the observations, and to the large-scale disc asymmetries.
The model presented in
KLSS will serve here as a reference case with which we compare the model
behaviour as we vary the parameters about its particular values.
In that model the computational length unit scaled to $\sim$270\au~with the
initial disc radius ($2$ units) and distance of pericentre ($2.6$ units)
for the perturber being $\sim$540\au~and $\sim$700\au~respectively.
Since $G$ and the primary mass were set to unity in the computations,
the computational time unit scaled to $\sim$530\yr~(i.e.
the period at unit radius divided by 2$\pi$).
Their system of computational units is used here in similar
numerical simulations,
however we point out that the scalings mentioned above
are only used for reference, in order to maintain generality
in the presentation of our results.

The new numerical calculations presented below employ a fifth-order
accurate Runge-Kutta-Fehlberg integrator with a variable time-step
size. While not as accurate as the Bulirsch-Stoer method
(e.g. as used by Hall et al. 1996), this method allows for the fast
integration of much larger numbers of particles ($\sim$10$^{4-5}$)
with moderate accuracy. Time-step size control is implemented as described in
Press et al. (1992), by specifying a maximum relative error
$\epsilon$ in the velocity magnitude. The value of $\epsilon$
was taken to be as small as practicable. Unless stated otherwise
we take $\epsilon\sim$10$^{-9}$. Thus, in a typical run of $\sim$10$^5$
time-steps, the net relative error in the energy and angular momentum
of any given particle is $\sim$10$^{-4}$. In practice the net relative
error for most of the particles will be substantially smaller than
this since the time-step size is controlled by a minority of particles
that fall close to either of the stellar masses.
To prevent too large a reduction in the time-step size owing to this effect
we remove particles from the calculation if they approach either star
to within a radius $R_{\rm p}$ of the primary or a radius $R_{\rm s}$
of the secondary. Unless stated otherwise we take: $R_{\rm p}=0.03$ and
$R_{\rm s}=0.01$. The former value was chosen as the minimum required to
ensure that the time-step size would not be controlled by close
approaches of particles to the primary rather than by close
approaches to the secondary, for all model parameters considered here.
The secondary was integrated as if it were a test particle in motion about a
fixed primary of mass equal to the total binary mass. The gravitational
forces on particles are then prescribed as a function of position at
each time-level with the addition of a fictitious force to compensate for the
non-inertial origin (see below). As mentioned above, the test
particles are collisionless and their distribution is
taken as a tracer of the observed dust presumed to be produced by
infrequent collisions in the real system.

The origin of time is taken as the instant of pericentre passage, hence
computational time-levels may be negative or positive. Throughout we
shall use $10^4$ test particles initialised in Keplerian
circular coplanar (zero
inclination) orbits, placed at random in the radial interval $0.5${--}$2$
such that the surface density profile decreases according to a
power law of radius to the three-halves. The perturber has inital
position and velocity components appropriate to the chosen orbital
inclination, eccentricity, and mass. The perturber is initialised
at a distance of ten times the initial outer
disc radius (i.e. at $20$ units). We shall consider a single
pericentre distance of $q=2.6$ units, thus the initial disc size is
$\sim$0.8$q$. The length and mass units, and therefore the time unit,
used in our calculations can be scaled so as to apply to any system
within the range of dimensionless parameters that we cover. Thus,
given our choice of relative disc size, fixing $q$ does not affect the
generality of our models.
Also, since the initial disc is axisymmetric, the longitude
of periastron for the perturber's orbit is arbitrary.
In all our particle position plots the perturber orbit is orientated
as in Fig. $3$, except for differences in orbital inclination.

\begin{figure}
\centerline{\epsfig{file=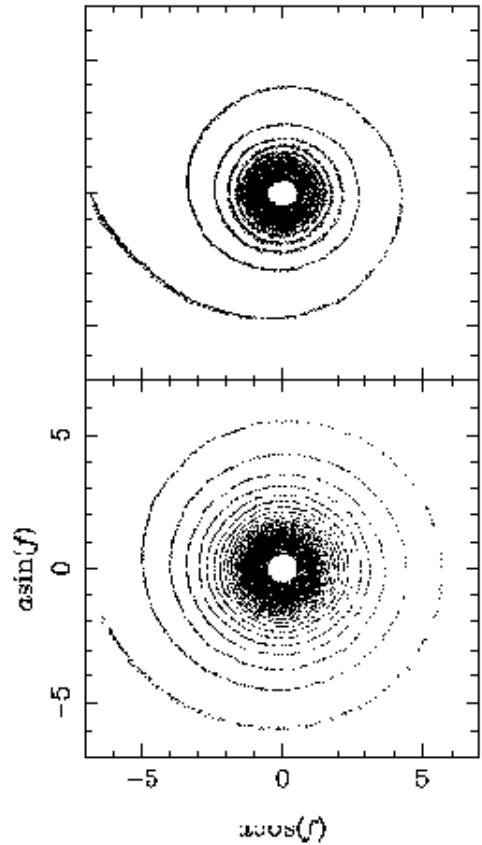,width=70mm,angle=0}}
\caption{Data for the model using only the indirect component of the tidal
potential (see Fig. $4$). The top frame is for data taken at a time of
$+60$; the bottom frame corresponds to a time of $+180$.
Plotted are $a\cos(f)$ against $a\sin(f)$ for each particle, where
$a$ is the semimajor axis and $f$ is the true anomaly.}
\label{fig5}
\end{figure}
\begin{figure}
\centerline{\epsfig{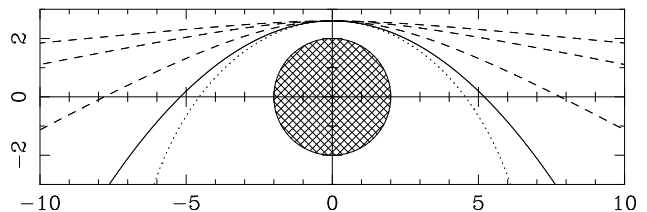}}
\caption{Schematic representation of the initial disc
(cross-hatched circle), and the orbital paths of a perturber
for a parabolic encounter (solid line) and hyperbolic encounters
(dashed lines, from lowermost to uppermost,
for $e=2$, $5$, and $10$ respectively). The dotted line corresponds to
a bound orbit with $e=0.75$ (see Section $3.5$).
As for the simulations the origin is assumed to be fixed to the
primary.}
\label{fig6}
\end{figure}

	\subsection{Circumstellar ring formation}

In order to identify the dynamical origin of the ring features
simulated by KLSS we consider the
system response to different components of the tidal potential
(an approach also used recently by Pfalzner \& Kley 2000, for gas discs).
The total potential is given by $\Psi =\Psi_{\rm p} + \Psi_{\rm s}$,
where the point mass potential at position ${\bf r}$ due to a
primary of mass $M_{\rm p}$, to which we have fixed the origin, is:

\begin{equation}
{\Psi_{\rm p}} = -\frac{GM_{\rm p}}{\mid {\bf r} \mid}.
\label{ppot}
\end{equation}

\noindent The tidal potential, due to the secondary with mass
$M_{\rm s}$ and instantaneous position ${\bf r}_{\rm s}$, is given by:

\begin{equation}
{\Psi_{\rm s}} = -\frac{GM_{{\rm s}}}{\mid {\bf r} - {\bf r}_{\rm s} \mid} 
+{GM_{{\rm s}}{\bf r} \cdot {\bf r}_{\rm s}\over |{\bf r}_{\rm s}|^3}.
\label{spot}
\end{equation}

\noindent The first term on the right-hand side of (\ref{spot})
is the direct potential due to the secondary star; the second
term is the indirect potential due to the acceleration of the origin.
In a stellar fly-by this acceleration corresponds to the reflex motion
of the primary star due to a gravitational tug from the secondary star.

In Fig. $4$ we give particle position plots for fly-by simulations
in which: we use only the direct term in the tidal potential,
use only the indirect term in the tidal potential, use the full tidal
potential.
In the case that employs only the direct tidal term the disc is
pulled towards the perturber as it approaches pericentre.
In the case using only the indirect tidal term the disc is
repelled by comparison as the perturber approaches pericentre
(in the inertial frame this effect corresponds to acceleration of the
primary towards the secondary).
In these first two cases eccentric nested rings are observed to form,
although they appear
as stronger features in the case where only the indirect tidal term
is used. Since the effectiveness of the direct tidal
interaction depends on the disc being able to closely approach the perturber,
the role of the indirect tidal term when the full tidal potential is used is
towards weakening the ring features that result from the direct tidal
interaction, and to cause approximate pericentric anti-alignment
of the two sets of rings.
This scheme in the alignment of the particle orbits gives a length asymmetry
between opposing sides of the disc when viewed edge-on,
with the strongest ring features along the longest extension.

In the above the features that appear to be ring-like
are actually formed from distended tightly-wound one-armed
kinematic spiral patterns. Different levels of advancement in the
winding process can be seen at the third and fourth time-levels
depicted in Fig. $4$.
This is more clearly illustrated in Fig. $5$ where we plot particle
positions according to the \emph{true anomaly} (i.e. the angle between
the instantaneous position vector and the line of pericentre)
for the last frame given in Fig. $4$ and for a time much later after
pericentre passage, for the model using only the indirect term in the
tidal potential. [Although we have now identified the origin of
the ring features, since over much of the azimuth the
appearence of the features resembles nested eccentric rings, we
shall continue with our original usage of that description.]

Winding of the kinematic spiral occurs as particles move around
their orbits with local orbital frequency $\Omega$.
Thus the time-scale for the
generation of a ring after pericentre passage
is $\Omega^{-1}$. The lack of significant width to the spiral
in Fig. $5$ as compared to Fig. $4$, indicates that the width
of the rings is mainly due to the spread in the particle
eccentricities. The rings will appear to propagate outwards
and dissolve due to the
growth of incoherence in the phase relationships of particles. Consider
two neighbouring particles at radii or semimajor axes
$a_{\rm o}$ and $a_{\rm i}$, where $a_{\rm o}>a_{\rm i}$.
The decoherence time-scale $t_{\rm d}$ for the particle pair is given by:

\begin{equation}
t_{\rm d}^{-1}= \frac{\Omega(a_{\rm i})}{2\pi}
\left[1-\left(\frac{a_{\rm i}}{a_{\rm o}}\right)^{3/2}\right].
\end{equation}

\noindent In $\beta$ Pic, and the simulations,
the width of the midplane brightness features is
$\sim$5--10 per cent of their radii (KLSS). For dissolution
of the ring features we require the tight winding of a spiral
down to the scale of a ring width, thus for a ring at
radius $r$:

\begin{equation}
t_{\rm d}\sim 10^5 \left(\frac{r}{500\au}\right)^{3/2}\yr.
\end{equation}

\noindent This agrees with the simulations of KLSS in which
it is found that the ring features dissolve in about ten local
orbital periods. KLSS also found that viewing the perturbed model disc at
approximately $10^5$\yr~after pericentre passage gave the most
satisfactory fit to the observed disc, which is consistent with
the innermost ring feature being detected at $\sim$500\au.

\begin{table}
\begin{center}
\begin{minipage}{63mm}
\caption{Data for some fly-by encounter models. The mass of the
secondary relative to the primary is $\mu$, the orbital eccentricity
of the secondary is $e$, and the orbital inclination of the secondary
is $i$. The length ratio of extension A to extension B is $L$.
We also give the percentage of the initial particle number that is ultimately:
bound to the secondary star ($n_{\rm b}$),
deleted by entering the secondary star's accretion radius ($n_{\rm s}$),
deleted by entering the primary star's accretion radius ($n_{\rm p}$), or
unbound from the system ($n_{\rm u}$). The model listed in the
third row was deemed to be most appropriate for modelling the
asymmetries in the $\beta$ Pic disc (KLSS, also Fig. $3$).}
\begin{tabular}{crcccccc}
$\mu$&$e$&$i$&$n_{\rm b}$&$n_{\rm s}$&$n_{\rm p}$&$n_{\rm u}$&$L$\\
\hline
0.1  &  1  & \phantom{1}0\degr & \phantom{1}6 & 2 & 0 & \phantom{1}3 & $\ll$1\\
0.3  &  1  & \phantom{1}0\degr &           12 & 5 & 0 & \phantom{1}4 & 1.5\\
0.3  &  1  &           30\degr &           11 & 0 & 0 & \phantom{1}5 & 1.2\\
0.3  &  1  &           60\degr & \phantom{1}0 & 0 & 0 & \phantom{1}0 & $\ll$1\\
0.3  &  2  & \phantom{1}0\degr & \phantom{1}9 & 0 & 1 &           12 & 0.3\\
0.3  &  2  &           30\degr & \phantom{1}8 & 0 & 0 & \phantom{1}8 & $\ll$1\\
0.3  &  5  & \phantom{1}0\degr & \phantom{1}0 & 0 & 0 & \phantom{1}7 & $\ll$1\\
0.3  & 10  & \phantom{1}0\degr & \phantom{1}0 & 0 & 0 & \phantom{1}1 & $\ll$1\\
0.5  &  1  & \phantom{1}0\degr &           16 & 1 & 6 & \phantom{1}5 & $\gg$1\\
1.0  &  1  & \phantom{1}0\degr &           21 & 2 & 8 &           17 & $\gg$1
\end{tabular}
\end{minipage}
\end{center}
\end{table}

\begin{figure}
\centerline{\epsfig{file=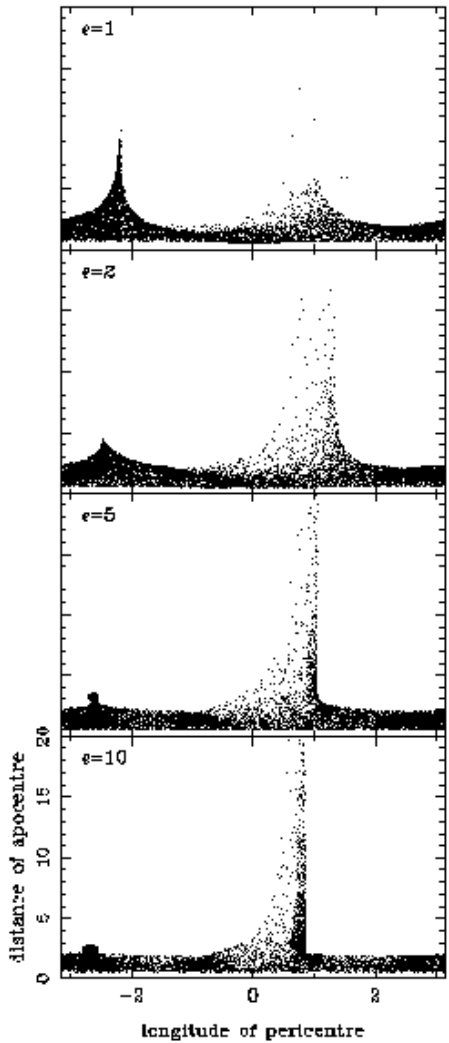,width=70mm}}
\caption{The final distance of apocentre against the longitude
of pericentre for fly-by simulations with perturber orbits of
eccentricity $e=1$, $2$, $5$, and $10$. Here and in similar plots we
select particles with $e<1$ and with separations from the secondary
of greater than 10, at the final time-level.
In all plots involving the longitude of pericentre,
zero longitude is defined by the positive horizontal
axis in face-on particle position plots.}
\label{fig7}
\end{figure}
\begin{figure}
\centerline{\epsfig{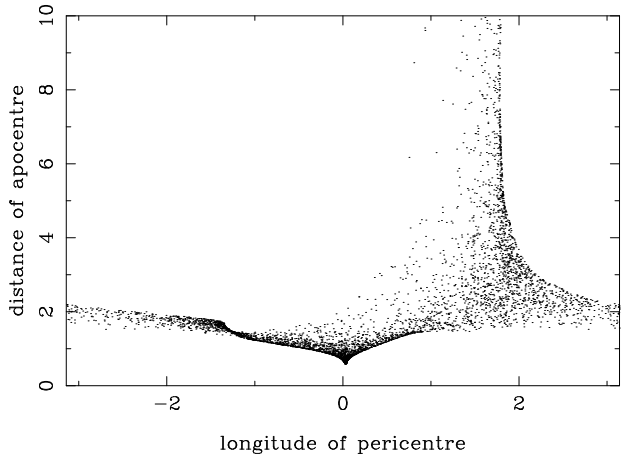}}
\caption{The final distance of apocentre against the longitude
of pericentre for a fly-by simulation with a parabolic perturber orbit
in which only the direct tidal potential has been used. Data
corresponds to those given in Fig. $4$.}
\label{fig8}
\end{figure}
\begin{figure}
\centerline{\epsfig{file=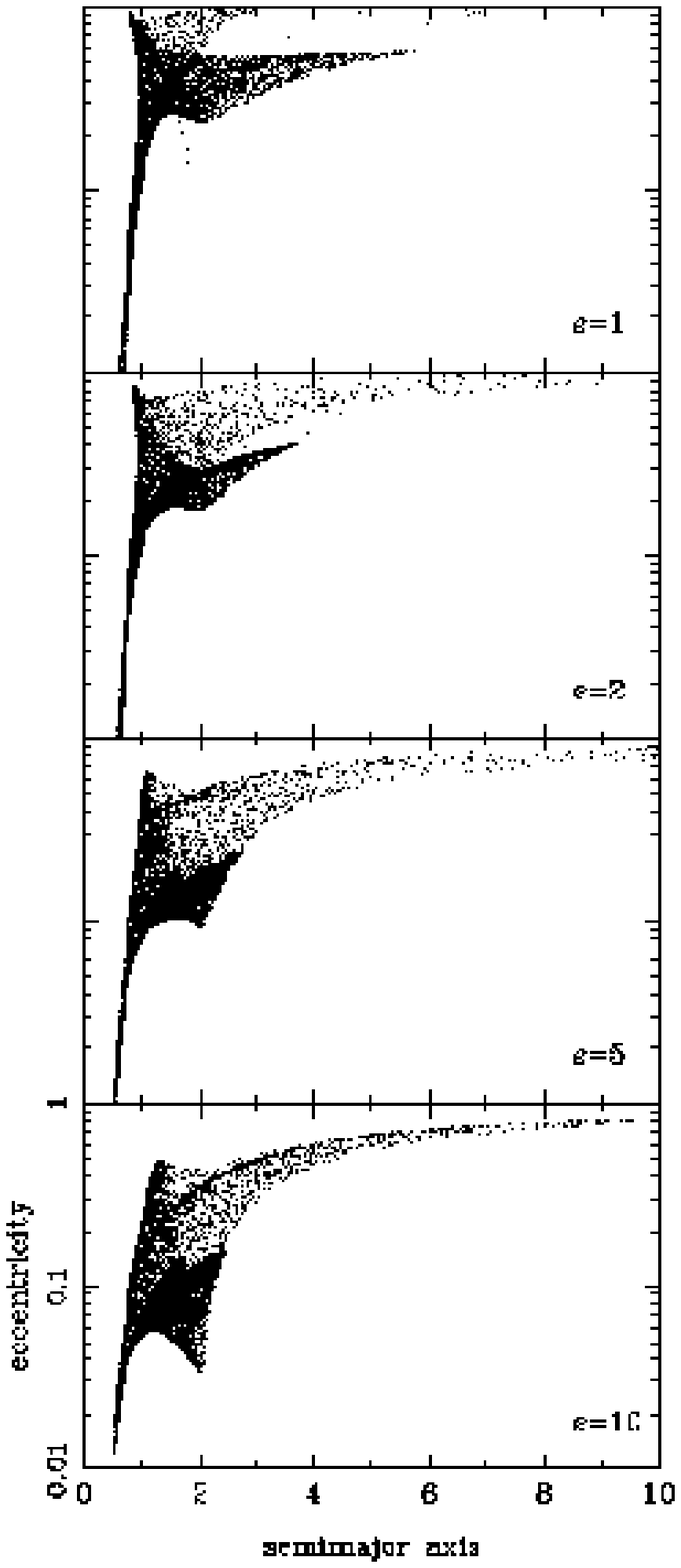,width=70mm}}
\caption{The final eccentricity against semimajor axis distribution
for fly-by simulations with perturber orbits of
eccentricity $e=1$, $2$, $5$, and $10$.}
\label{fig9}
\end{figure}

	\subsection{Length asymmetry in coplanar models}

As discussed above, a length asymmetry in the disc is readily obtained
from a close fly-by encounter. The length asymmetry in $\beta$ Pic's disc is
also the most easily measured of its asymmetric features. Below we shall
continue to numerically investigate the form and size of the length
asymmetry as a function of fly-by parameters.

		\subsubsection{Parabolic versus hyperbolic trajectories}

A secondary star of orbital eccentricity $e$ with respect to
the primary star has relative velocity
magnitude $V_{\rm q}$ at pericentre distance $q$ given by:

\begin{equation}
V_{\rm q}^2 = (1+e)\frac{G(M_{\rm p} + M_{\rm s})}{q}.
\label{vq}
\end{equation}

\noindent Hence for the scaling used by KLSS,
eccentricities of $1$, $2$, $5$, and $10$ correspond to pericentre
velocities of $\sim$2.5, $3.1$, $4.3$, and $5.9\,$km\,s$^{-1}$ respectively.
In this Section we shall consider coplanar encounters of uniform
pericentre distance $2.6$, a perturber mass of $0.3$, and these
orbital eccentricities. The corresponding trajectories
and the initial disc size used in these simulations are shown in Fig. $6$.

A useful diagnostic measure of the length asymmetry induced in the
disc is given by plotting the final apocentre distance against
the longitude of pericentre for each perturbed particle orbit.
By \emph{final} we shall mean that the secondary mass has passed
through pericentre and has returned to a sufficiently large radius that
its affect on the positions and velocities of the particles bound
to the primary is negligible.
In our simulations we integrated the system for a total time
of $120$ units, which means that the final perturber distance was
always much greater than the initial distance, corresponding
to a time of at least $80$ units after pericentre passage.
This total run-time
was also generally required in order for the numbers of particles
either bound or unbound to each of the two stellar masses to stabilise (in
practice this meant to approach a constant value with less than
about $10$ per cent fluctuations; cf. Hall et al. 1996). This
information is summarised in Table $1$, for all the $e\ge1$ models
presented in this paper.

In Fig. $7$ we compare the disc response for encounters of different
eccentricity. In all cases there are two oppositely directed extensions.
The extension near to a longitude of $-2$ contains the
strong ring features, and the other extension (near longitude $+1$)
contains the weak ring features, as discussed above.
Respectively we shall refer to these as extension A and extension B.
Thus extension A forms as a response to the indirect
component of the tidal potential, and extension B forms
as a response to the direct component of the tidal potential.

As the encounter velocity (equivalently, the orbital eccentricity)
increases the peak of extension A diminishes in amplitude, while
the peak of extension B increases its amplitude.
Further examination of the extension B results illuminates a key difference
between the parabolic ($e=1$) and hyperbolic ($e>1$) orbit cases.
In the former, extension B is compact, but in
the latter cases all show a much longer extension B.
In Fig. $8$ we show the final apocentre distance against the longitude
of pericentre for the parabolic case in which only the direct part
of the tidal potential is used. As we might have expected
extension A is not in evidence (being due to the indirect part of
the tidal potential), and extension B is similar in nature
to that found in the hyperbolic orbit cases.
Thus we conclude that in the full tidal potential case
the indirect component acts
on the disc prior to pericentre passage such that the disc
response due to the direct component near pericentre is
modified as compared with
a calculation that uses only the direct term
(also discussed in Section $3.1$). Hence the more
impulsive (i.e. hyperbolic) encounters yield a response that is
similar to that found for the parabolic case that uses the
direct term only, being manifest in the form of extension B.
It follows that higher velocity (i.e. more eccentric)
perturbers will have extension B forms more like that found
in Fig. $8$, and smaller amplitudes for extension A. This
can be seen in Fig. $7$.

In Fig. $9$ we give the final eccentricities of the disc particles
for different eccentricities.
In the parabolic orbit case, extension A particles correspond to
the wing-shaped distribution having eccentricities approximately
in the range $0.2${--}$0.6$. Superposed with this are the extension
B particles that have generally very high eccentricites. The change
in the relative lengths of extensions A and B with increasing
perturber eccentricity is also apparant. Generally, the excited
eccentricities of the disc particles are lower for larger
perturber eccentricities, which can be explained from the simplistic
impulse model for tidal interaction (Lin \& Papaloizou 1979):
higher relative velocity results in weaker coupling between
the particles and the perturber.

\begin{figure}
\centerline{\epsfig{file=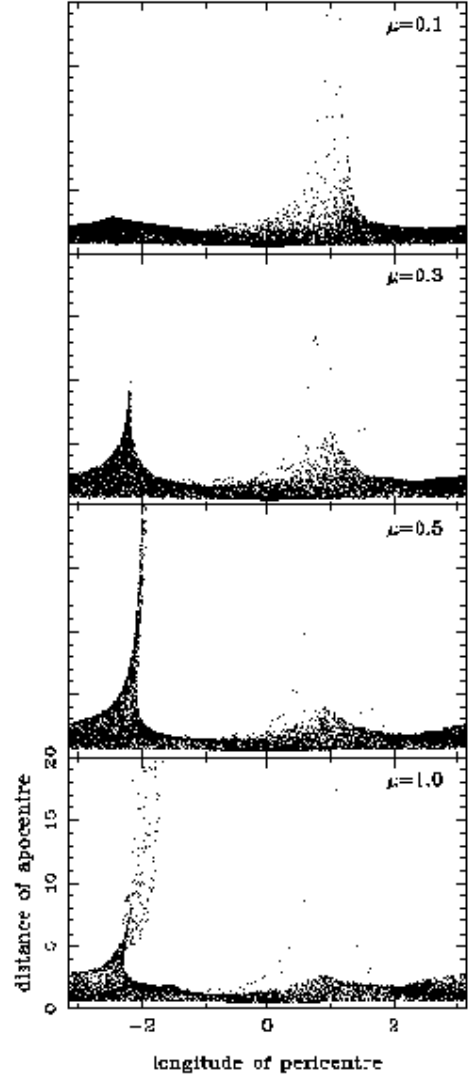,width=70mm}}
\caption{The final apocentre distance against longitude of pericentre
distribution for parabolic fly-by simulations with perturber masses of
$0.1$, $0.3$, $0.5$, and $1.0$.}
\label{fig10}
\end{figure}
\begin{figure}
\centerline{\epsfig{file=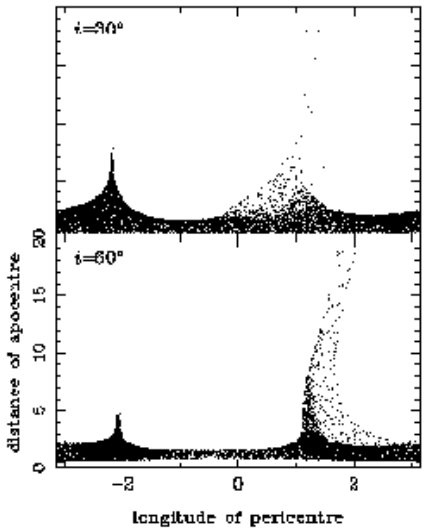,width=70mm}}
\caption{The final distance of apocentre against the longitude
of pericentre for parabolic fly-by simulations with orbital
inclination $i=30$\degr~and $60$\degr.}
\label{fig11}
\end{figure}

		\subsubsection{Variation of the perturber mass}

In Fig. $10$ we show the final distance of apocentre against the final
longitude of pericentre for parabolic encounters with different masses for
the perturber. We consider the mass ratio $\mu = M_{\rm s}/M_{\rm p}$.
Increasing the perturber mass from $\mu =0.1$ to $\mu =1.0$,
we find that extension A increases in amplitude and
extension B diminishes in amplitude. The former effect is due to
the larger perturber mass inducing a larger amplitude reflex motion
of the primary star; the latter effect evidently results from the
more massive perturber being able to strip more particles from
the disc in the direct interaction that sets up extension B
(see Section 3.4).

The lowest mass model with $\mu=0.1$ has an extension B that has
a similar extended form to the high eccentricity cases discussed
above. This is due to the low mass of the perturber giving only
a small reflex motion of the primary, and so a relatively weak interaction
prior to pericentre passage, that necessarily results in a more impulsive
response. Only the $\mu=0.3$ model gives a compact extension B and
extension A.
This case has in terms of a model for the
length asymmetry in the disc of $\beta$ Pic the most appropriate
amplitude ratio for the extensions A and B of those coplanar
models tried here.
The ratio of lengths of apocentre for this model is approximately
$9/6=1.5$, the length ratios of the other models clearly being
very different from unity. The length ratio between extensions A
and B is given in Table $1$, for all our models with $e\ge1$.
Where the ratio is clearly much different from unity we simply
state either $\gg1$ or $\ll1$, since the interpretation of
the length of an extension as defined by a small number of particles
at large distances is not clear.
 
We note that the size of extension A can be reduced
by increasing the perturber's eccentricity, so it may be possible that
a more massive ($\mu>0.3$) and eccentric orbit ($e\ga1$)
perturber could fit the
observational results. However, since eccentricities larger than
approximately unity do not display a compact extension B, we favour
the perturber mass originally considered by KLSS (i.e. $\mu\sim0.3$,
for $e\sim1$).

\begin{figure*}
\centerline{\epsfig{file=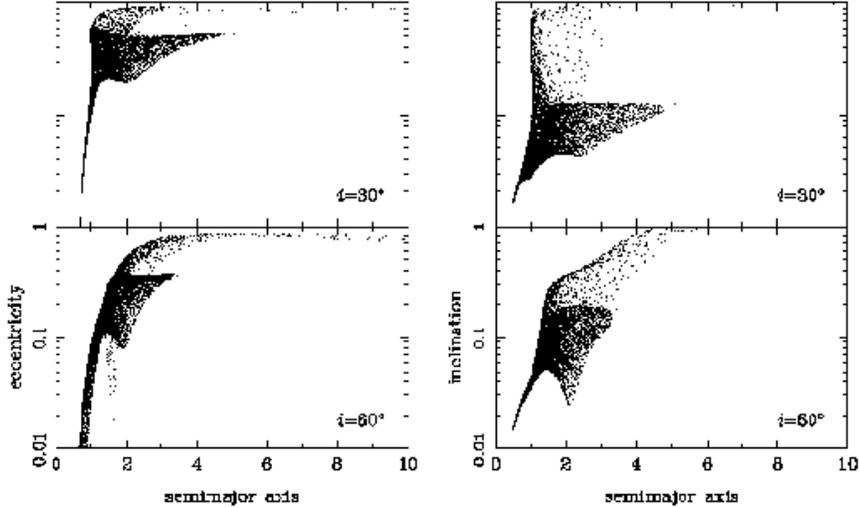,width=120mm}}
\caption{The final eccentricity and inclination against semimajor axis
distributions for parabolic fly-by simulations with orbital
inclination $i=30$\degr~and $60$\degr.}
\label{fig12}
\end{figure*}

	\subsection{Inclined orbits}

In general, if the perturber's orbit and the initial disc plane are
not coplanar, then the tidal interaction occurring
between the disc particles and the perturber can excite
vertical motions in the particle orbits. Here we shall
consider this effect in terms of the inclination $i$ of the
perturber's orbit to the initial disc plane. This angle is generated
by rotating the perturber's orbit about the \emph{latus rectum}
(i.e. the line perpendicular to the line of pericentre, that also passes
through the origin), which
then becomes the \emph{line of nodes} (e.g. see Fig. $3$).

KLSS found that (for $i=30$\degr) extension B became vertically flared,
with much weaker inclinations in extension A,
as can be seen in Fig. $3$. From the preceeding discussion we
deduce that the flaring results from the stronger direct tidal
interaction, and the smaller inclinations generated in extension A
result from the weaker indirect tidal interaction.
The resulting morphology resembles the
butterfly asymmetry (Kalas \& Jewitt 1995, KLSS) in which the vertical
height of the imaged disc is greatest north of the midplane in the
southwest extension (see Figs. $1$ and $2$),
and south of the midplane in the northeast extension,
with the overall greatest vertical height in the former.

		\subsubsection{Length asymmetry}

In Fig. $11$ we give the final distance of apocentre against the
longitude of pericentre for parabolic encounters with $\mu=0.3$,
and $i=30$\degr~and $i= 60$\degr. In the lower inclination
case the disc response is very
similar to the coplanar encounter presented above, except that
extension A is shorter by about $20$ per cent. In the
higher inclination case the response resembles that of a coplanar
hyperbolic encounter, having a very short extension A and a very
long extension B. As before, this is consistent with the notion that
the disc interaction prior to pericentre passage is a key difference
between hyperbolic and parabolic encounters. In the $i=60$\degr~case
the high inclination takes the perturber orbit further from
the disc, making the pericentre interaction more impulsive. This
shows that we require an orbit with not too high an inclination,
if we wish to preserve the compact nature of extension B.
Additionally, the other angle possible, which corresponds to rotation
of the orbit about the line of pericentre, need not be considered
since for low values of $i$ that angle takes the orbit yet further
from the disc. In other words, we only consider an
argument of pericentre of $90$\degr.

In the higher inclination case the length ratio of extension A
to extension B is clearly much less than unity. In the lower
inclination case extension B is compact, but the peak is less
distinct than for the coplanar case. We measure its height by
taking the intersection of the tangents to the sides of the
envelope of the extension particles near its base. This yields a height of
approximately $6$, which is similar to the coplanar model.
Thus we deduce a ratio $7/6\sim1.2$. This is the best fitting
value, in terms of the length asymmetry, of those models presented
in this paper. Scaling according to the KLSS model
the lengths of the extensions are $1890$\au~and $1620$\au, which
is in good agreement with the new determination of the size
of the disc presented in Section $2$.

\begin{figure}
\centerline{\epsfig{file=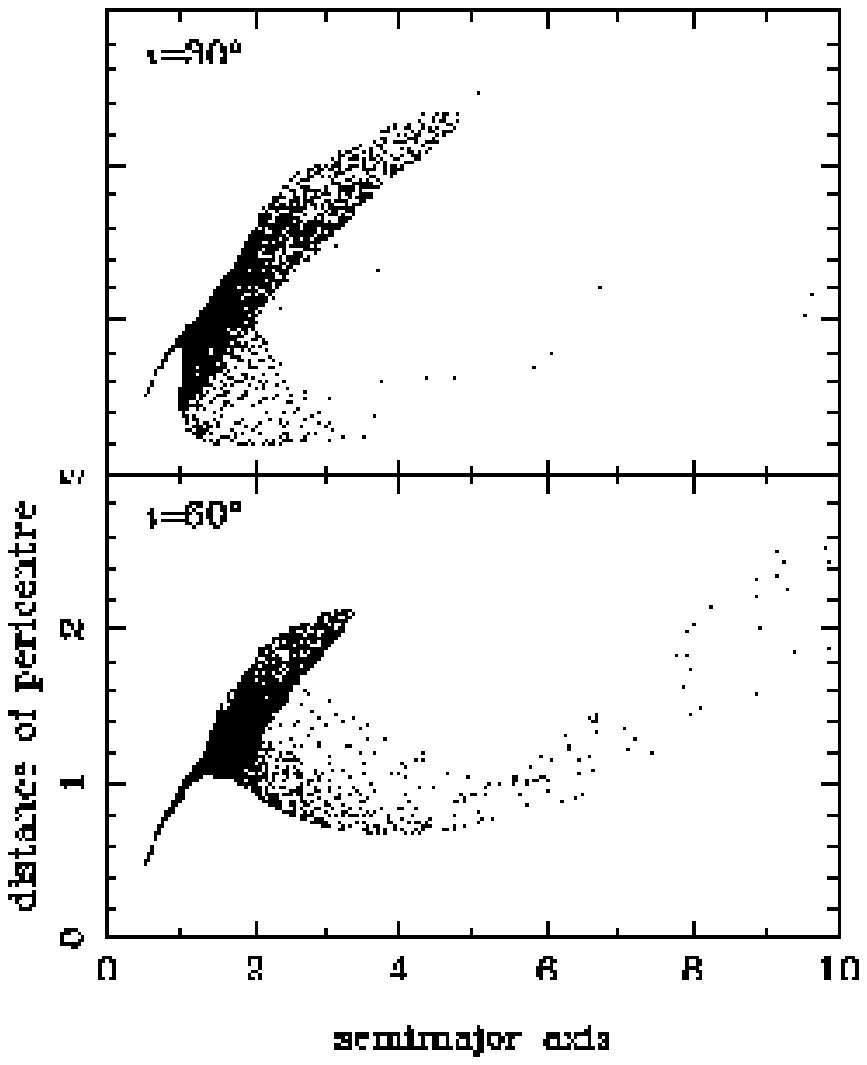,width=70mm}}
\caption{The final distance of pericentre against semimajor
axis for parabolic fly-by simulations with orbital
inclination $i=30$\degr~and $60$\degr.}
\label{fig13}
\end{figure}
\begin{figure*}
\centerline{\epsfig{file=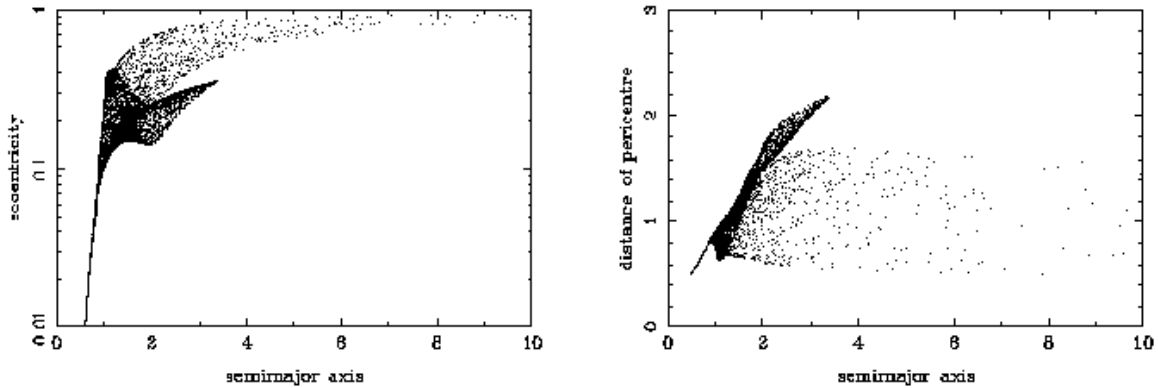,width=60mm,angle=270}}
\caption{The final eccentricity and distance of pericentre against semimajor
axis for a hyperbolic fly-by simulation with orbital eccentricity
$e=2$ and orbital inclination $i=30$\degr.}
\label{fig14}
\end{figure*}

The exact numerical value of the length ratio determined for the
$\beta$ Pic disc
could be recovered in our models by taking a slightly smaller orbital
inclination than $30$\degr.
That would not be productive at this stage, however, since here we
are not concerned with the details of translating our results into
a scattered light disc model, which could introduce a small
correction factor. However, for reference, a model with $i=15$\degr
was tried and was found to give a length ratio of $1.4$. Thus
we expect the appropriate inclination to lie between $15$\degr~and
$30$\degr, i.e. closer to the lower inclination model above than to coplanar.

		\subsubsection{Flaring and warps}

In Fig. $12$ we show the final eccentricity and inclination
distributions for the disc particles in the two cases mentioned above.
The amplitude of the response is lower for higher values of $i$,
for the reasons indicated above.
The eccentricity response resembles that of a coplanar hyperbolic orbit,
as expected from the preceeding discussion. The inclinations associated
with extension A are much smaller than for those associated with
extension B. However, the extension A inclinations are higher in the
$i=60$\degr~model, owing to input from the direct interaction
because of increased impulsiveness of the interaction occuring at
the ascending node in that case. Thus, by symmetry of the orbit
with respect to the initial
disc, we expect the inclinations in each extension to
become comparable when orthogonality is approached by increasing $i$.
This provides a further need for a low inclination encounter
as the disc of $\beta$ Pic is strongly lopsided with respect to
vertical flaring (Kalas \& Jewitt 1995).

For these two cases we also give the final pericentre distance
distributions in Fig. $13$. In the low inclination case we see that
particles from extension B can plunge to pericentre distances of
$0.2$. KLSS proposed that these particles could
explain the existence of an inner 'warp' in the scattered light
dust disc of $\beta$ Pic at $\sim$50\au~(Burrows et al. 1995,
Mouillet et al. 1997). We can see from Fig. $13$,
that for the higher inclination encounter, extension B particles
have significantly higher perturbed pericentre distances (down to about $0.7$;
or $\sim$200\au). This is a direct consequence of the
lower eccentricities produced at comparable semimajor axes
in extension B for the two cases. Thus the minimum pericentre
distance is minimised in coplanar cases.
Correspondingly, larger minimum pericentre distances are also
charateristic of hyperbolic encounters.
Fig. $14$ gives the final eccentricity
and pericentre distributions for a hyperbolic encounter with
$e=2$ and $i=30$\degr. If the explanation of the inner warp
suggested by KLSS is confirmed then this also supports the need for a
low inclination and relative velocity encounter.

	\subsection{Planetesimal capture}

As can be seen in Fig. $4$ the close fly-by of a disc of particles
by a perturbing star can result in the very close approach of some of the disc
particles to the perturber. Since the unperturbed disc does not extend
all the way out to pericentre, this occurs by particles streaming through
the vicinity of the instantaneous Lagrange Point between the two stars,
about the time of closest approach. In the case that particles
become captured by the perturber the capture efficiency is
generally low (see also Hall et al. 1996).
Table $1$ gives the percentage of initial disc
particles that are captured by the perturber for the models discussed above,
and from it we can see that the lower eccentricity and inclination
encounters are required, in order to effect capture at all.

For the model parameters used by KLSS, the percentage of captured particles
is $\sim$10 per cent of the initial particle number,
but note that if the initial
disc were allowed to extend to pericentre this figure could be fractionally
larger (Hall et al. 1996). In the case of $\beta$ Pic, considering
the maximal parent body disc of Backman \& Paresce (1993), the captured
material could therefore represent up to about $10^3$ earth masses
of planetesimals which would subsequently orbit the perturber and
produce a scattered light signature similar to Vega-like sources.
This would not however uniquely identify a perturber candidate
since approximately 15 per cent of main sequence stars of all types
are known to have such discs (Backman \& Gillett 1987, Aumann 1988),
but given the encounter parameters we find here, it seems likely that
a perturber candidate should have this signature.

None the less we may find crude approximations for the orbital
properties of the captured planetesimals. We shall make an analogy
with the theory of mass transfer in semi-detached binaries,
in which the captured matter orbits the secondary with the
specific angular momentum of relative motion at the
Lagrange Point (see Frank, King \& Raine 1992). Here we deduce that if
a particle in the outermost parts of the disc is captured close to
the perturber's pericentre, then it will orbit the perturber with
specific angular momentum:

\begin{equation}
\left[
\sqrt{\frac{G(1+e)(M_{\rm p}+M_{\rm s})}{q}} -
\sqrt{\frac{GM_{\rm p}}{R_{\rm o}}}
\right]
b,
\end{equation}

\noindent where $R_{\rm o}$ is the initial outer radius of the
disc and $b$ is the separation of the secondary from the
Lagrange Point at the instant of pericentre passage,
approximated by:

\begin{equation}
\frac{b}{q}= \frac{1}{2}+0.227\log (\mu).
\end{equation}

\noindent In the above we have taken the circular velocity of a
particle lying on the binary axis relative to the perturber
at pericentre for the input particle velocity.
Although the instantaneous Lagrange Point is
located well within the envelope of the initial disc, the rapid
eccentricity changes in the disc prior to particle transfer
cause compression of the disc such that the outer disc particles
pass through the vicinity of the Lagrange Point with approximately their
initial velocities (e.g. see Fig. $4$).

\begin{figure*}
\centerline{\epsfig{file=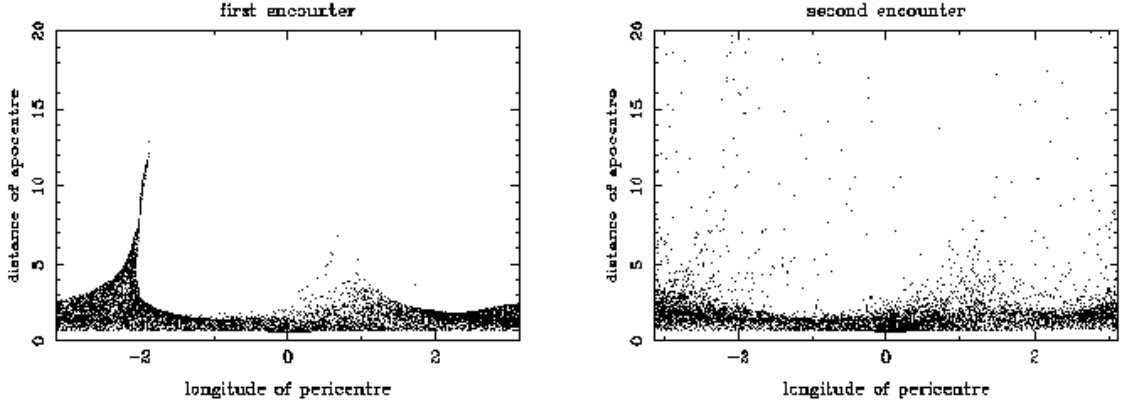,width=60mm,angle=270}}
\caption{The final distance of apocentre against the longitude
of pericentre for a bound orbit fly-by simulation with $e=0.8$.
The companion starts at apocentre and was set up and integrated much as
for the hyperbolic orbit cases. Data is given for the
first and second complete orbits. In this simulation we used
$\epsilon=10^{-7}$.}
\label{fig16}
\end{figure*}

With respect to the perturber, the specific angular momentum
of the captured particle is:
$\sqrt{GM_{\rm s}a'(1-e'^2)}$, where
the characteristic semimajor axis and eccentricity with respect
to the secondary are respectively $a'$ and $e'$.
Hence we have:

\begin{eqnarray}
a'(1-e'^2) &=&
q(1+e)\frac{(1+\mu)}{\mu}\left(\frac{b}{q}\right)^2\nonumber\\
&&\times\left[ {1-\sqrt{\frac{q/R_{\rm o}}{(1+e)(1+\mu)}}} \right]^2.
\end{eqnarray}

\noindent Thus for a model with $e=1$, $\mu=0.3$,
and $q=2.6$, we find $a'(1-e'^2)\sim0.3$.
This equation does not apply when the relative
velocity of the perturber and disc particles is sufficiently
high that particle transfer is energetically disallowed.
As shown in Table $1$ we find that particles are
captured when $e\le2$ but not when $e\ge5$.

The above is in fair agreement with the simulations. In Fig. $15$
we give the positions, eccentricities, and semimajor axes of captured
particles relative to the perturber for the above model.
We find the bound particles have a median value $a'\sim0.70$,
from which we calculate $e'\sim0.76$. The actual median value is
$e'\sim0.85$ for the particles, which is $\sim10$ per cent larger
than predicted.
This is likely due to the continued interaction of the primary and the
captured particles as the perturber receeds from pericentre. 
The particle transfer does not occur in a well defined stream
which results in a large spread in semimajor axes and eccentricities.
However, the captured particles appear to manifest as an asymmetrical
disc about the perturber (upper panel of Fig. $15$), and show
clustering in $e'$ and $a'$ (lower panel of Fig. $15$),
indicating some level of coherence in the transfer (seen in Fig. $4$).
Scaling according to KLSS, we find that $90$\% of the disc particles
are located within $500$\au~of the perturber. For the
KLSS model (i.e. the previous model augmented with an orbital
inclination of $30$\degr~for the perturber), the median eccentricity
and semimajor axis are each found to be $\sim 0.7$,
consistent with the higher relative disc velocity expected
for the inclined encounter.

The luminosity of an M0V dwarf star (of mass $\sim 0.5$\,M$_\odot$,
corresponding to $\mu=0.3$) is $\sim$10$^2$ times smaller than
for $\beta$ Pic. If the amount of scattering surface in the disc
is proportional to the square of the number of parent bodies
(e.g. Backman \& Paresce 1993),
then as an upper limit the brightness
of the perturber's disc is expected to be $\sim$10$^4$ times
fainter than that around $\beta$ Pic, and $\sim$10$^2$ times fainter
relative to the central star. Thus it is plausible that
the perturber may demonstrate a measurable scattered light disc
as a result of collisional erosion amongst the captured planetesimals
at the distance to $\beta$ Pic.

\begin{figure}
\centerline{\epsfig{file=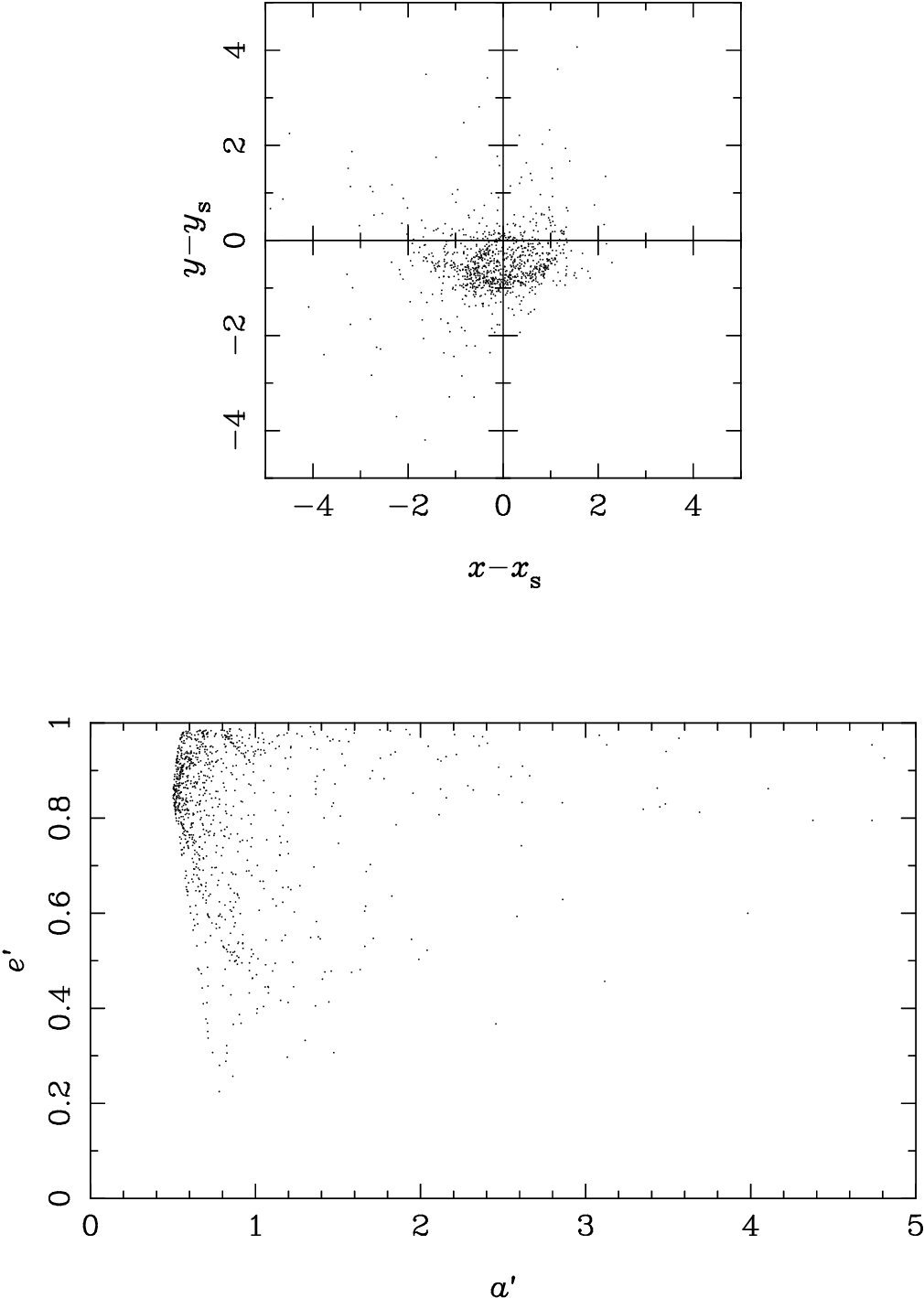,width=80mm,angle=0}}
\caption{Particle positions and eccentricity versus semimajor
axis for captured particles relative to the perturber.}
\label{fig15}
\end{figure}

	\subsection{Close encounters with a bound perturber}

Since the innermost midplane brightness feature identified by KLSS
is measured to be at $\sim$500\au~the
age of the rings in the $\beta$ Pic disc should be
$\sim$10$^5$\yr, as calculated above.
Although this figure could be revised up or down
according to future observations, we shall take this to mean that
the epoch of the fly-by encounter was about $10^5$\yr~in the past.
Use of the \emph{Hipparcos} survey data reveals that this is
also the minimum time-scale
we must look back from the present in order to find a
perturber fly-by at $\la 10^5$\au~(Kalas et al. 2000).
The eighteen most closely approaching stars identified by Kalas et al. (2000)
have a mean encounter velocity $\sim$40\kmps, which is
considerably higher than that deduced from our dynamical calculations.
Thus we are prompted to consider companions that are or were physically
associated with $\beta$ Pic.

Repeated close encounters of the bound star with the disc
may destroy the asymmetrical disc structure and ring features set up
by the first encounter. Since for $e<1$ the semimajor axis of the perturber
can be written $q/(1-e)$, by demanding the orbital period to
be greater than some value $P_{\rm min}$, we deduce:

\begin{equation}
e>1-q\left[
\frac{4\pi^2}{G(M_{\rm p}+M_{\rm s})P_{\rm min}^2}
\right]^{1/3}.
\end{equation}

\noindent Thus if we take $P_{\rm min}=10^5$\yr~we infer that the
perturbed binary eccentricity needs to be greater than $0.75$,
in order that we may view the ring features at their inferred age.

In Fig. $6$ we show a trajectory with $e=0.75$.
The similarity of such high eccentricty bound orbits to
the parabolic trajectory suggests that the disc response should be
similar. In Fig. $16$ we give the apocentre distance versus longitude of
pericentre plots for a coplanar encounter simulation with $e=0.8$,
$\mu=0.3$, and $q=2.6$. The perturber starts
at apocentre and we give results for the first and second completed
orbital periods. Clearly the length asymmetry is completely disrupted after the
second encounter. In Fig. $17$ we give the corresponding particle
positions, which show that the ring structure is also destroyed after
the second encounter. Thus as suspected only a single fly-by can be allowed.

Possible scenarios in which a single pericentre passage could have
occurred in the recent past are:
1) the perturber was originally a bound companion to $\beta$ Pic with
a sufficiently large pericentre distance that no significant
interaction with the disc occurred over its lifetime, and sufficently
large semimajor axis and eccentricity that it could be perturbed into the
required trajectory by the close fly-by of a massive field star
$\sim$10$^5$\yr~in the past;
2) $\beta$ Pic was a hierarchical multiple
system that was dynamically unstable on a time-scale comparable to its
present age, resulting in the ejection of
one or more stars and the development of a disc-intercepting orbit for
another, or an ejected star passed by $\beta$ Pic on its way out of the system.

In the first scenario the three-body interaction
should have occurred at least $10^5$\yr~ago. The further in the past
we put this event, the more eccentric the resultant binary orbit must
be in order to delay closest approach until $10^5$\yr~ago (if the
perturbed star has an initially out-going trajectory). Although perhaps
more attractive than the lower probability close binary encounter
model used by KLSS, the problem of low relative velocity is shifted from the
star-disc interaction to the
interaction between the second and third bodies.
In respect of the second scenario we point out that
Weinberger et al. (2000) report an age of $5$\myr~for the triple
system HD 141569. This consists of a Vega-like BVe primary with a
$500$\au~disc and two M stars each at $\sim$1000\au~projected separation.
Weinberger et al. (2000) deduce that the relative separations involved
may make the system unstable, giving the possibility that HD 141569
represents an earlier stage in the evolution of the $\beta$ Pic system
(with an age of $20$\myr).
Studies of the dynamics involved suggest that young stars with
multiple companions may not lose some of them via dynamical
instability until main sequence ages (Eggleton \& Kiseleva 1995),
which is supported by observations of high binary frequency amongst
young stars compared to field stars (Mathieu 1994).

\begin{figure}
\centerline{\epsfig{file=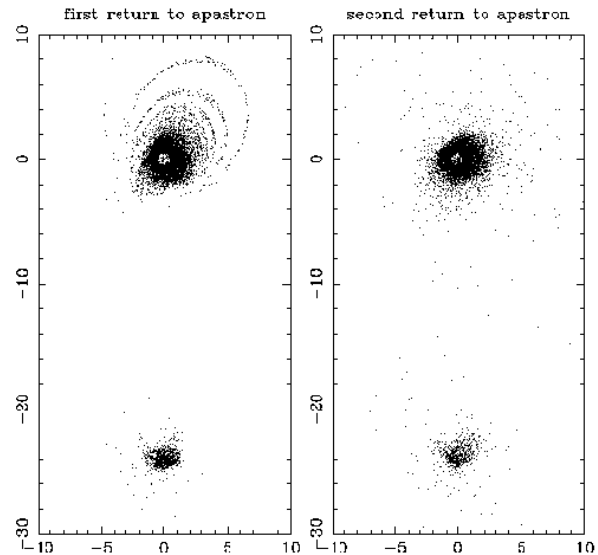,width=80mm,angle=270}}
\caption{Particle positions corresponding to data given in Fig.$16$.
The primary's position is plotted with an asterisk.}
\label{fig17}
\end{figure}
\begin{figure}
\centerline{\epsfig{file=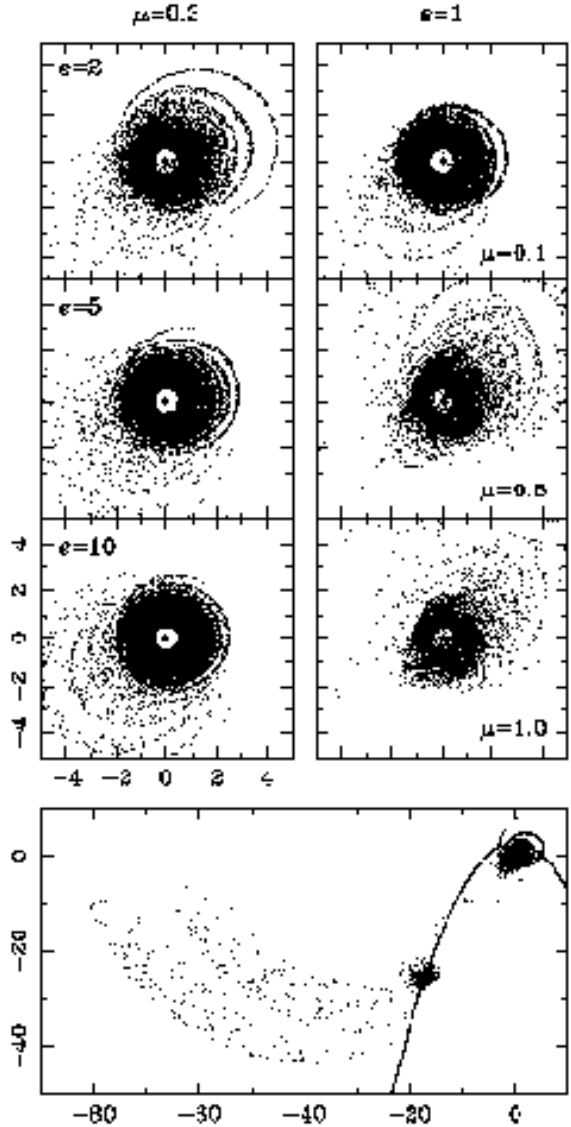,width=85mm,angle=0}}
\caption{Summary of coplanar model results in face-on particle
plots. The models were run for $120$ time units, with the perturber
initialised a radial distance of $20$ units from the primary.
After much longer
times the phase-mixing in the particles' orbits results in
dissolution of the ring features, maintaining the distended envelopes
of the discs seen above.
The position of the primary is plotted with an asterisk.
The top left-hand side panel shows results for models
with $\mu=0.3$ and from top to bottom $e=2$, $5$, and $10$ respectively.
The top right-hand side panel shows results for models
with $e=1$ and from top to bottom $\mu=0.1$, $0.5$, and $1$ respectively.
The bottom panel shows results, on a larger scale, for the model
with $e=1$ and $\mu=0.3$. The perturber trajectory is shown with a solid
line. We see: the perturbed disc with rings, the captured particles
clustered about the instantaneous position of the perturber, and
escaping particles (the large arcs on the left-hand side of the plot).}
\label{fig18}
\end{figure}
\begin{figure}
\centerline{\epsfig{file=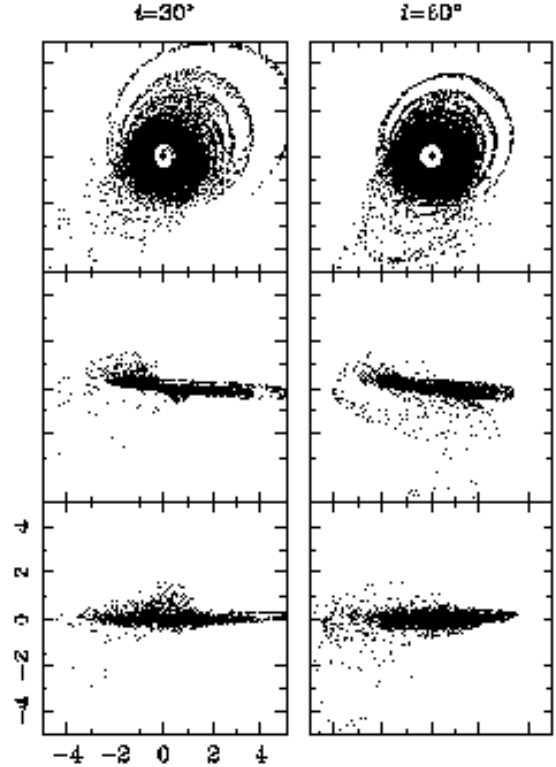,width=85mm,angle=0}}
\caption{Summary of inclined-orbit models in face-on and
edge-on particle plots. The models were run for $120$ time units,
with the perturber initialised a radial distance of $20$ from the primary.
The left-hand panel shows results for
the model with $e=1$, $\mu=0.3$, and $i=30$\degr. The right-hand
panel shows results for the corresponding case with $i=60$\degr.
Uppermost frames are face-on views, middle frames are edge-on
projections along the horizontal axis (line of nodes), the bottom
frames are edge-on projections along the axis defining $i=0$.
Thus the projections are on to the three mutually orthogonal Cartesian
planes of the computational coordinate system.
The position of the primary is plotted with an asterisk.}
\label{fig19}
\end{figure}

\section{Discussion}

Here we have considered the stellar fly-by hypothesis that may
account for the asymmetrical structure in the scattered light
dust disc of $\beta$ Pic. We investigated the dynamics of the
perturbed planetesimal disc in an attempt to learn more about
the possible parameters of a fly-by encounter. Additionally,
we have described the origin of transient circumstellar eccentric ring
structures as being a general outcome of an encounter, due to the reflex motion
of the primary star as the perturber passes through closest approach.
The apparent ring system is actually an eccentric tightly-wound one-armed
spiral pattern that gradually disappears owing to phase-mixing in the
particles' orbital motion. In inclined-orbit simulations we find
that the vertical disc response also has analogues in the scattered light
disc of $\beta$ Pic. We summarise the coplanar results in Fig. $18$,
and the inclined-orbit results in Fig. $19$.

For the perturber we conclude that the mass should not be very
different from $0.5$\,M$_\odot$ (corresponding to spectral type M0V).
In this case we favour a low inclination prograde encounter with
relative velocity of a few km\,s$^{-1}$, which makes a bound
perturber seem the most natural choice. One problem with this is
that asymmetries would be destroyed in the second pericentre passage.
Thus, given the deduced youth of the asymmetries, the perturber
in this scenario is required to make a single close approach in the
recent past (i.e. $\sim$10$^5$\yr~ago). This can be accounted for
by making the perturber-$\beta$ Pic system an initial wide binary that
is perturbed by a massive passing field star $\sim$10$^5$\yr~in the
past, such that the perturber is sent into a close approach orbit about
$\beta$ Pic. A more probable solution is to arrange for $\beta$ Pic
to have originally been a hierachical multiple system (cf. HD 141569;
Weinberger et al. 2000) that became dynamically unstable on the age
of the star, and with the result being that one component was sent
into a disc-intercepting orbit. In any case the perturber should
have captured planetesimals from the primary's disc and may exhibit
a detectable scattered light disc as a result.
A survey of a more extended field around $\beta$ Pic than has been
considered previously is in progress in order to test our hypotheses.

The dynamical response of the models suggests that the $\beta$ Pic
disc might be separable into different components corresponding
to groupings of perturbed particle orbits. For example, we find that outwards
of $\sim$250\au~the SW extension may not have a distinct midplane
owing to pumping of particle inclinations. The positions close to
pericentre of the moderately eccentric NE midplane particles could account
for the measurement of a SW midplane out to $\sim$650\au, in
superposition with the more diffuse vertically and radially extended component.
When the optical image is vertically compressed (Fig. $2$)
we are able to measure the
SW extension to much larger radii ($1450$\au). In our model the
corresponding (very eccentric
and inclined) particle orbits also intersect the disc near to their
pericentres (down to $\sim$50\au) possibly accounting for the inferred
inner warp of the disc, which is aligned with the outer flared
envelope in the SW. In summary, there are three recognizable particle
groupings that can be related to the morphology of the $\beta$ Pic disc.
These are: the highly eccentric and inclined particles that
reach apocentre in the SW (extension B particles),
the moderately eccentric and weakly
inclined particles that reach apocentre in the NE (extension A particles),
and the relatively unperturbed particles inside $\sim$200\au~radius
(Figs. $12$ and $13$).

\section*{Acknowledgments}
We are grateful to the University of Hawaii for supporting the observations.
This work was funded in part by a grant from NASA held by D. Jewitt.
We also thank J. Gradie,
B. Zuckerman, and E. Becklin for the use of their coronagraph.
The clarity of this paper was improved by comments from S. Ida.
P. Kalas acknowledges financial support and hospitality given by the
Astronomy Unit, Queen Mary \& Westfield College, during his visit there.

\label{lastpage} 
\end{document}